\documentclass[conference]{IEEEtran}
\IEEEoverridecommandlockouts
\usepackage{graphicx}
\usepackage{amsmath,amssymb,amsfonts}
\usepackage{algorithmic}
\usepackage{textcomp}
\usepackage{xcolor}
\usepackage{url}

\def\BibTeX{{\rm B\kern-.05em{\sc i\kern-.025em b}\kern-.08em
    T\kern-.1667em\lower.7ex\hbox{E}\kern-.125emX}}
    
\usepackage[numbers, sort]{natbib}

\begin{document}

\title{Navigating the Data Space Landscape: Concepts, Applications, and Future Directions}

\makeatletter
\newcommand{\linebreakand}{%
  \end{@IEEEauthorhalign}
  \hfill\mbox{}\par
  \mbox{}\hfill\begin{@IEEEauthorhalign}
}
\makeatother

% *\\

\author{
\IEEEauthorblockN{Bojana Marojevikj}
\IEEEauthorblockA{\textit{Faculty of Computer Science and Engineering} \\
\textit{University Ss Cyril and Methodius}\\
Skopje, N. Macedonia\\
bojana.marojevikj@finki.ukim.mk}
\and
\IEEEauthorblockN{Riste Stojanov}
\IEEEauthorblockA{\textit{Faculty of Computer Science and Engineering} \\
\textit{University Ss Cyril and Methodius}\\
Skopje, N. Macedonia\\
riste.stojanov@finki.ukim.mk}
}

\maketitle

\begin{abstract}
This paper explores the evolving landscape of data spaces, focusing on key concepts, practical applications, and emerging future directions. It begins by introducing the foundational principles that underpin data space architectures, emphasizing their importance in facilitating secure and efficient data exchange. The core design principles and essential building blocks that form the backbone of data-space systems are then examined. Several real-world implementations are presented, showcasing how data spaces are applied across various industries to address challenges such as data sovereignty, interoperability, and trust. The paper concludes by discussing future directions, proposing that leveraging semantic data models can significantly enhance interoperability and data integration within data spaces. Furthermore, it suggests exploring the implementation of SPARQL as a sophisticated authorization layer to improve security and granular control over data access. This research provides a comprehensive understanding of the current state of data spaces and aims to guide future advancements in this rapidly evolving field by highlighting the potential of semantic data and SPARQL-based authorization.\end{abstract}
\begin{IEEEkeywords}
Data spaces,  International Data Spaces Association, Eclipse Dataspace Connector, Reference Architecture Model, Gaia-X, SPARQL\end{IEEEkeywords}

\section{Introduction}
In an era characterized by exponential data growth and the increasing need for seamless data exchange, the concept of data spaces has emerged as a potential solution. The purpose of data spaces is to break down data silos and enable trusted data sharing, fostering innovation across ecosystems. They create a common ground where multiple stakeholders, such as businesses, governments, research institutions, and even individual users, can securely share and collaborate on data while maintaining control over who can access what. This approach not only boosts efficiency but also unlocks new opportunities for collaboration and value creation.
For example, in healthcare, a data space can allow hospitals, research laboratories, and pharmaceutical companies to share anonymized patient data for research purposes. This could accelerate drug discovery, improve personalized medicine, and improve patient outcomes, while complying with strict privacy regulations such as GDPR \cite{voigt2017eu}.
Data spaces rely on technologies such as identity management, access control, and data usage policies to ensure secure interactions. Initiatives like Catena-X \cite{CatenaX} in Europe illustrate how these ecosystems work in practice, promoting transparency, interoperability, and digital sovereignty.
This paper will explore the core concepts of data spaces, highlight real-world applications, discuss challenges, and outline future directions. By understanding how data spaces function and their transformative potential, we can better navigate the evolving digital landscape and unlock new frontiers of innovation.

\section{Core Concepts}

The concept of data spaces emerged around 20 years ago within the field of computer science as an alternative to data sharing through restrictive bilateral agreements between organizations \cite{franklin2005databases}. Unlike centralized methods, such as data consolidation hubs, data spaces do not require physically merging data into a single location. Instead, data remains stored at its original source. Additionally, a unified database schema is not necessary for integrating data from multiple sources. Instead, integration occurs at the semantic level through shared vocabularies. This flexibility allows for data redundancies and the coexistence of different datasets. Data spaces can also overlap or nest within each other, allowing participants to belong to multiple data spaces at the same time \cite{otto2022evolution}. One of the first real implementations of the concept of data space emerged in 2015 with the International Data Space (IDS) initiative \cite{ids_2025}.

\subsection{Design Principles}
The core design principles that guide the development and operation of data spaces, ensuring that they are secure, fair, and sustainable are the following:
\begin{itemize}
    \item \textbf{Data Sovereignty:}  
    Data sovereignty refers to the right of individuals and organizations to independently control their economic data assets. This is a core principle in data spaces, empowering participants to: (1) manage, process, and secure their data more effectively, and (2) retain control over data access when sharing with others. \cite{nagel2021design}

    \item \textbf{Level Playing Field for Data Sharing and Exchange:}  
    Creating a fair environment within data spaces ensures that all participants, including new entrants, can compete based on data quality and service offerings rather than data volume. This fairness is supported by promoting collaboration over competition, achieved through careful design and ongoing maintenance of the underlying infrastructure. \cite{nagel2021design}

    \item \textbf{Decentralized Soft Infrastructure:}  
    The infrastructure for data spaces is supposed to be decentralized and composed of interoperable systems adhering to shared functional, technical, operational, and legal standards. Although invisible to participants, this "soft infrastructure" should handle interoperability, security, privacy, and trust through advanced identity and consent management. It should also support data monetization, while remaining technology neutral to allow participants flexibility in choosing their tools and platforms. \cite{nagel2021design}

    \item \textbf{Public-Private Governance:}  
    Effective governance is crucial for building and sustaining data spaces. It requires representation from businesses, individuals, governments, and technology partners to balance public and private interests. Existing regulations (such as GDPR \cite{voigt2017eu} and eIDAS \cite{sharif2022eidas}) provide a foundation, and new policies are being developed to further support data spaces. The public sector will initially drive adoption through funding and participation, with long-term success driven by network effects and community growth. \cite{nagel2021design}
\end{itemize}

\subsection{Building blocks}
The architectural landscape of data spaces is characterized by several recurring patterns that reflect the core principles of the concept. A prominent pattern is that of a decentralized architecture, where data remains under the control of the originating producers, and access to these data is granted in a carefully managed and secure manner. This approach stands in contrast to centralized data repositories, such as data lakes. Another significant pattern is the federated architecture, which involves the interconnection of multiple data spaces based on a foundation of shared policies and rules. This federation enables data sharing across diverse industries, companies, and entities, creating a network of interconnected data stores rather than a single monolithic repository. Additionally, some initiatives, such as the International Data Spaces Association (IDSA) \cite{ids_2025}, propose the IDSA Reference Architecture Model (RAM) \cite{ids_ram_2025}), a conceptual model which serves as a blueprint with the purpose of providing common architectural principles and guidelines for organizations to create or join a Data Space. It may be considered as the current leading architecture reference for Data Space initiatives and can be seen as a set of technical components within the broader context of building blocks as shown in Figure 1.

\begin{figure}[ht]
  \centering
  \includegraphics[width=\columnwidth]{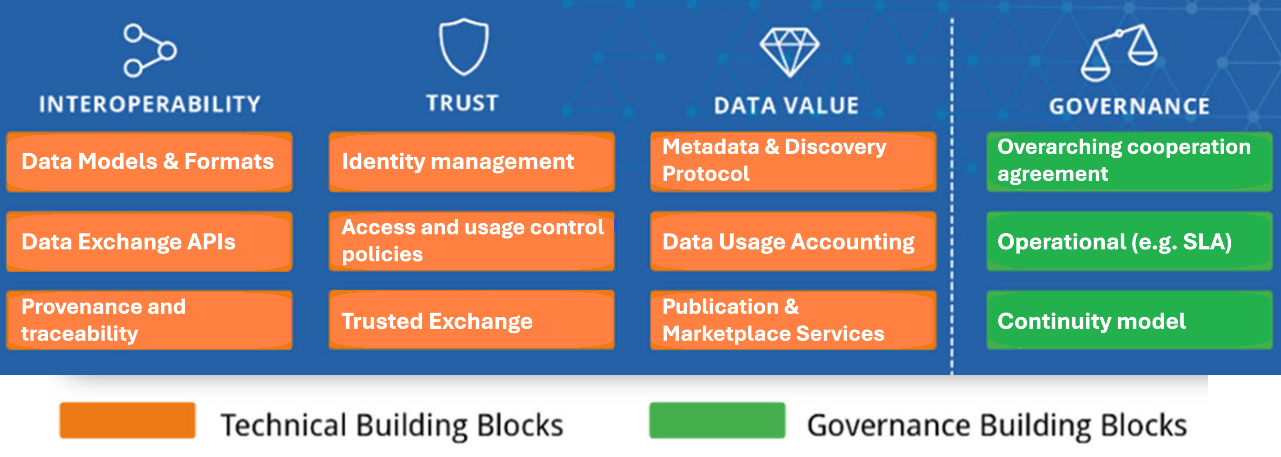}
  \caption{An overview of technical and governance building blocks for data spaces \cite{nagel2022build}}
  \label{fig:building-blocks}
\end{figure}

Several essential technical components are integral to the functioning of data spaces. As illustrated in Figure 2, data sharing in a data space involves several key technical components working in coordination. \textbf{Connectors} serve as the critical technical interface, facilitating data exchange by implementing and enforcing the policies established within the data space. Connectors manage the negotiation of contracts between data providers and consumers and ensure adherence to the usage rules defined by participants. They also enable secure data transfer through a variety of protocols and application programming interfaces (APIs), such as Java Database Connectivity (JDBC) \cite{sciore2020jdbc}, Amazon Simple Storage Service (Amazon S3) \cite{aws_s3}, and Hypertext Transfer Protocol Secure (HTTPS) \cite{durumeric2013analysis}. Examples of connector technologies include Eclipse Dataspace Connector (EDC) \cite{eclipse_edc_connector}, FIWARE TRUE Connector \cite{fiware_true_connector}, and Simpl \cite{eu_simpl}. The effectiveness of data spaces depends on these connectors, which provide the technical means for secure and policy-compliant data sharing across distributed systems.

\textbf{Data catalogs} and \textbf{discovery services} play a vital role in enabling participants to register and locate available data assets and services within the data space ecosystem. These components store metadata pertaining to data offerings, including crucial details such as usage policies and access conditions. They also support functionalities for searching and browsing the array of available data sources. Efficient data discovery is crucial for data spaces, enabling participants to find required data while adhering to providers' policies.

\textbf{Identity and Access Management (IAM)} systems are fundamental to ensuring a secure data sharing environment by managing identities and authentication process. This often involves the use of digital identities and verifiable credentials to establish trust among participants and facilitate secure transactions. Robust IAM capabilities are essential for building confidence and ensuring that only authorized entities can access and utilize data within defined policy parameters.

Policy enforcement mechanisms are critical for managing the agreements between data providers and consumers prior to any data exchange. These mechanisms ensure that agreed-upon data usage policies are strictly enforced, thereby guaranteeing compliance with the conditions stipulated by data owners. This may involve sophisticated contract negotiation processes and the application of policy languages, such as the Open Digital Rights Language (ODRL) \cite{ianella2007open}. Automation of policy enforcement is vital for maintaining data sovereignty and ensuring that data are used in accordance with the agreed terms, thus minimizing the need for manual oversight.

Finally, shared services encompass a range of functionalities that support the operation and governance of the data space. These can include services for participant registration and validation, certification processes, monitoring of data space activities, and overall data space management functions. The management and provision of these shared services can be centralized or decentralized, depending on the specific requirements and design of the data space. Shared services are crucial for ensuring a level playing field for all participants and maintaining the overall integrity and functionality of the data space ecosystem.

\begin{figure}[ht]
  \centering
  \includegraphics[width=\columnwidth]{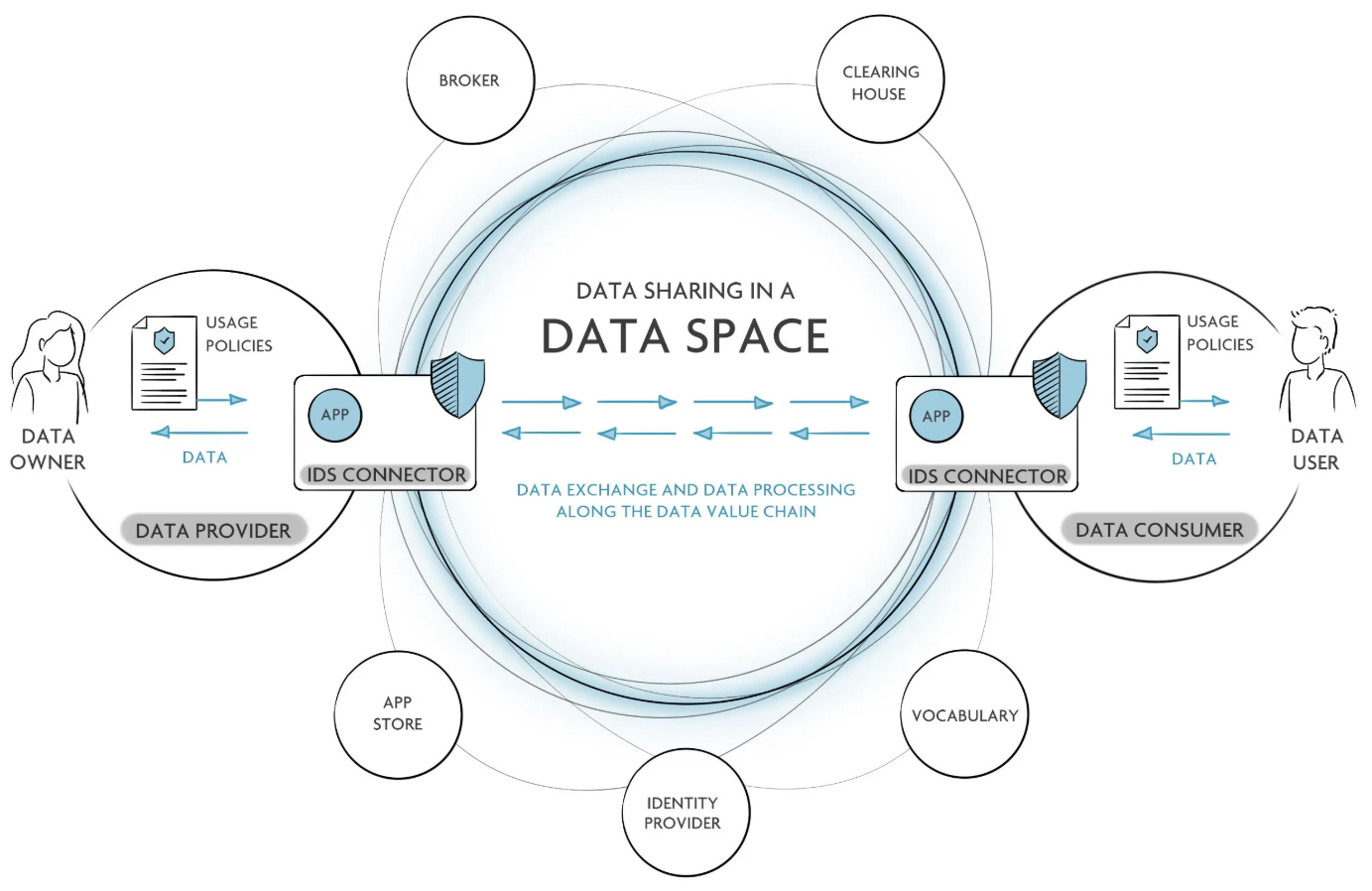}
  \caption{Data sharing in a data space \cite{IDSA2025}}
  \label{fig:building-blocks}
\end{figure}

\section{Existing Technical Developments and Initiatives}
The data space landscape is being actively shaped by several key organizations and initiatives that are driving standardization, providing frameworks, and fostering collaboration. The International Data Spaces Association (IDSA) \cite{ids_2025} is an association dedicated to creating standards for data sharing within data spaces, thereby empowering participants with complete control over their data assets. IDSA Reference Architecture Model (RAM) \cite{ids_ram_2025} is a crucial framework for the design and implementation of data spaces. The association's core objectives revolve around promoting data sovereignty and ensuring interoperability across diverse data ecosystems, underpinned by governance principles emphasizing accountability, transparency, fairness, and responsibility. 

Gaia-X \cite{tardieu2022role} represents a significant European initiative aimed at developing a federated and secure data infrastructure grounded in European values of transparency, openness, data protection and security. With the main goal of ensuring European digital sovereignty and fostering the digital economy, Gaia-X focuses on building a network of trust among industrial partners. It provides a comprehensive blueprint for the creation of secure, transparent, and interoperable data infrastructures. 

The Data Spaces Support Centre (DSSC) \cite{dssc} plays a coordinating and supportive role in the development of data spaces throughout Europe, with funding from the European Commission. DSSC is responsible for developing the Data Space Blueprint \cite{dssc_blueprint}, which outlines essential building blocks and design principles for data spaces. The center provides a variety of tools and support services to address the organizational and technical requirements of establishing and operating data spaces. Its primary aim is to establish common requirements and promote best practices to accelerate the creation of sovereign data spaces within the European digital landscape.

The Data Spaces Business Alliance (DSBA) \cite{dsba} represents a collaborative effort between IDSA \cite{ids_2025}, the Big Data Value Association (BDVA) \cite{bdva}, Gaia-X \cite{tardieu2022role}, and the FIWARE Foundation \cite{fiware2025}. The primary focus of this alliance is on achieving technical convergence among existing data space architectures and models, with the goal of ensuring interoperability and portability across different data space implementations. DSBA represents a concerted effort to harmonize the diverse initiatives in the data space domain, fostering greater cohesion and interoperability within the ecosystem. 

Several key open-source projects and platforms are also contributing significantly to the technical foundation of data spaces. The Eclipse Dataspace Components (EDC) \cite{eclipse_edc} project provides an open source framework for building secure and globally scalable data-sharing services. EDC offers a set of customizable components for constructing control planes, data planes, decentralized identity systems, and federated data catalogs. It is built on the specifications of the Gaia-X Trust Framework \cite{tardieu2022role} and the IDSA Dataspace protocol \cite{ids_dataspace_protocol}  and is utilized by initiatives such as Eona-X \cite{eona_x} and Catena-X \cite{CatenaX}. EDC serves as a fundamental open source resource, providing the essential building blocks for implementing data spaces based on established standards and frameworks.  

FIWARE \cite{fiware2025} offers a range of open source components that can be used to build data spaces. In particular, the FIWARE TRUE Connector offers a dedicated specification to enable secure and efficient data sharing within the IDS ecosystem. FIWARE's contribution lies in providing reusable open-source tools that facilitate the development of interoperable data space solutions.

Simpl \cite{eu_simpl} is an open source, smart and secure middleware platform initiated by the European Commission to facilitate the creation of EU data spaces and enhance interoperability between them. Simpl aims to streamline data sharing across various data spaces and enable data scaling from the edge to the cloud. Its initial deployment is focused on key public sector domains such as healthcare, public procurement, and mobility.

\section{Real-World Implementations}
The potential of data spaces is being realized in various industries through concrete implementations. In the logistics industry, the Smart Freight Centre (SFC) Exchange Network \cite{AWSDataSpaces} is a collaborative network built as a data space on AWS. Its primary goal is to promote transparency and decarbonization in transport chains by facilitating the exchange and reporting of emissions data from logistics. 

The automotive industry has seen the development of Catena-X \cite{CatenaX}, a prominent data space driven by more than 60 companies. This initiative aims to address challenges and opportunities in areas such as traceability, sustainability, the circular economy, and efficient supply chains throughout the automotive value chain. Catena-X heavily relies on the Eclipse Dataspace Connector (EDC) as a core technology for enabling secure and sovereign data exchange among its participants.

The healthcare sector is another area where data spaces are making significant strides. The European Health Data Space (EHDS) \cite{NTTData2025} aims to create a secure and interoperable environment for the exchange of health data, to accelerate medical research and improve healthcare delivery. Similarly, Dataspace4Health \cite{NTTData2025}, with contributions from NTT DATA Luxembourg, provides a secure environment for health data exchange, enabling new medical research while protecting patient privacy.

In the realm of mobility and smart cities, the Mobility Data Space (MDS) \cite{NTTData2025} serves as a platform for aggregating, managing, and sharing mobility-related data from various sources. This facilitates data collaboration for purposes such as traffic management, urban planning, and transportation optimization. Another data space initiative in this sector is the Flanders Smart Data Space \cite{desemantic} which aims to establish
a comprehensive and interoperable data infrastructure in the region of
Flanders (Belgium). This data space leverages a combination of key technologies to achieve its objectives. The Eclipse Dataspace Connector (EDC) is utilized to provide secure and sovereign data exchange capabilities among participants. Complementing this, Semantic Web technologies are employed to ensure semantic interoperability across diverse datasets. The core technology for data dissemination within the space is Linked Data Event Streams (LDES), which enables near real-time access and synchronization of information for data consumers.

Another initiative is the Green Deal Data Space (GDDS) \cite{lush2024assessing} which is designed to facilitate the secure and interoperable exchange of high-quality data between sectors to support sustainability and resilience goals. It integrates diverse datasets related to climate change, biodiversity, zero pollution, circular economy, smart mobility, and environmental compliance. Built on FAIR principles \cite{wilkinson2016fair}, the GDDS aims to overcome fragmented data systems by connecting public, private, and citizen-generated data within a trusted digital ecosystem. 

In the energy sector, the development of data spaces is being supported through various EU-funded initiatives. The Energy Data Space is a key focus of the ETIP SNET policy framework \cite{etip2023}, aiming to enable secure, standardized, and interoperable data exchange across energy systems and actors. Coordinated by the BRIDGE initiative \cite{bridge2021}, efforts in this domain have led to significant advancements in cross-dataspace interoperability. Notable testbeds include the OMEGA-X \cite{omegaXWebsite} and ENERSHARE \cite{enershareWebsite} data spaces, which have demonstrated practical scenarios for interoperable and sovereign data sharing in energy applications. These initiatives not only address technical challenges but also contribute to the establishment of a federated ecosystem for energy data exchange aligned with European digital and energy policy goals \cite{omega2025standardization}.

\section{Future Directions}
The current landscape of data spaces, while promising, faces several critical challenges that threaten its widespread adoption and the realization of a truly interconnected data economy. One of the primary challenges lies in achieving seamless interoperability and semantic understanding across diverse data sources. Although existing standards and protocols provide a foundation, a robust semantic layer is crucial for enabling machines to interpret and process data accurately. This necessitates the development of comprehensive ontologies and knowledge graphs that can capture the context and meaning of data across various domains. Furthermore, ensuring secure and granular access control within data spaces remains a significant hurdle. Current authorization mechanisms often lack the flexibility and precision needed to manage complex data sharing scenarios. To address this, future directions should explore the integration of advanced access control techniques, such as the SPARQL authorization guard. This approach aims to create a more intelligent and secure data environment where: 
\begin{itemize}
    \item Data are annotated with semantic metadata, enabling automated discovery, integration, and reasoning. This will facilitate cross-domain data exchange and create a richer, more contextualized data landscape.
    \item A SPARQL authorization guard will enable fine-grained control over data access based on semantic queries and policies. This will allow data providers to define precise rules for who can access what data and under what conditions, enhancing security and trust.
    \item The combination of semantic interoperability and robust access control mechanisms will contribute to a more dynamic data economy. This enables data providers to monetize their assets, while allowing data consumers to securely access relevant information.
\end{itemize}
\begin{figure}[ht]
  \centering
  \includegraphics[width=\columnwidth]{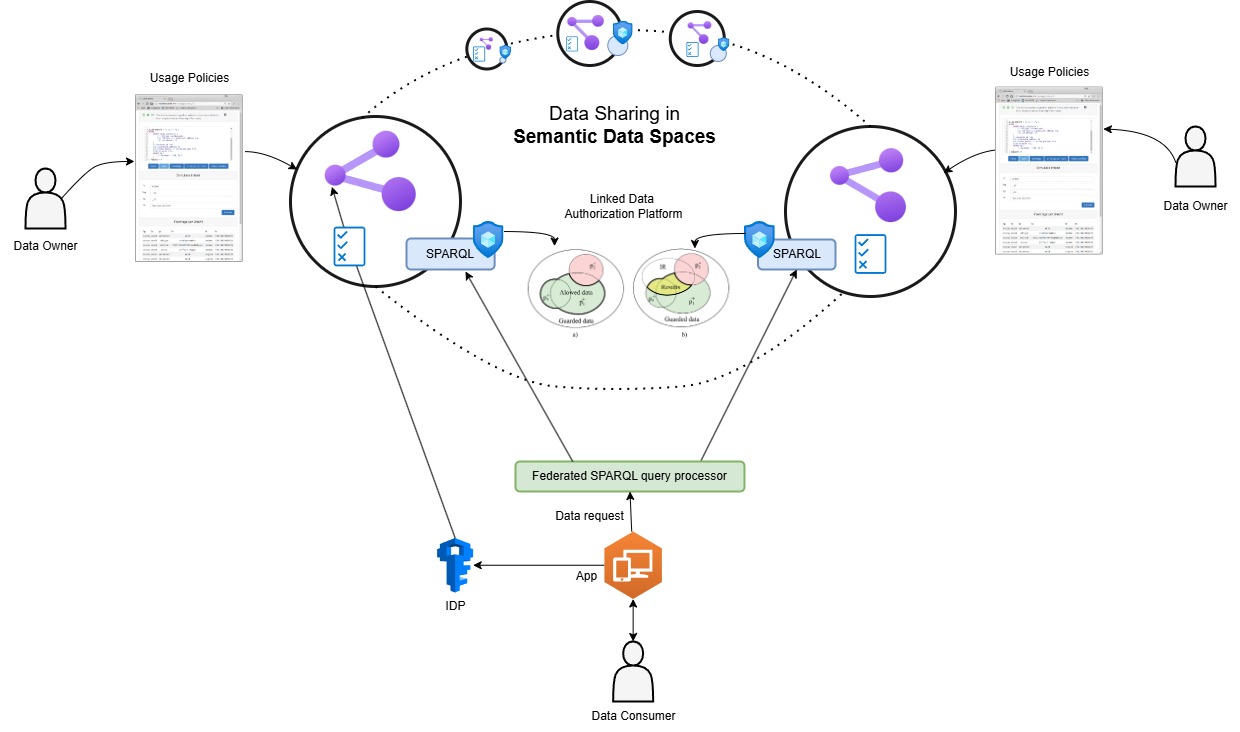}
  \caption{Semantic data space with an authorization layer}
  \label{fig:building-blocks}
\end{figure}

\section{Conclusion}
This research paper has explored the complex landscape of data spaces, exploring their key definitions, principles, wide-ranging applications, and potential future developments. Data spaces represent a paradigm shift in how data is managed and shared, moving away from traditional centralized systems toward decentralized, collaborative ecosystems. The key principles of data sovereignty, interoperability, trust, security, and governance underpin this novel approach, addressing critical concerns associated with data exchange in an increasingly interconnected world.

The applications of data spaces are vast and varied, spanning industries from logistics and automotive to healthcare and public services. Real-world implementations demonstrate their potential to facilitate data integration, enhance collaboration, strengthen governance, and power data marketplaces, ultimately unlocking significant value for individuals, organizations, and society. The benefits are multifold, ranging from increased control over personal data and improved public services to fostering innovation and driving economic growth.

However, the journey towards widespread adoption of data spaces is not without its challenges. Technical complexities, data quality issues, integration hurdles, and concerns about privacy and security must be carefully addressed. The lack of standardization remains a significant challenge, highlighting the need for collaborative efforts to establish common frameworks and protocols.

Looking ahead, the future of data spaces is set to evolve with a greater emphasis on semantic interoperability, granular access control, and secure data exchange. The development of comprehensive ontologies and knowledge graphs will enhance the ability to interpret and integrate data across domains, fostering a more interconnected ecosystem. At the same time, fine-grained access control mechanisms, such as SPARQL authorization guards, will enable precise and policy-driven data sharing, strengthening security and trust. As data spaces continue to mature, ongoing research and collaboration among stakeholders will be essential in overcoming existing challenges and unlocking their full potential for a more efficient and dynamic data economy.

\bibliographystyle{ieeetr}
\bibliography{refrerences}

\begin{thebibliography}{10}

\bibitem{voigt2017eu}
P.~Voigt and A.~Von~dem Bussche, ``The eu general data protection regulation (gdpr),'' {\em A practical guide, 1st ed., Cham: Springer International Publishing}, vol.~10, no.~3152676, pp.~10--5555, 2017.

\bibitem{CatenaX}
{Catena-X}, ``{Catena-X: The Automotive Network}.'' \url{https://catena-x.net/en/1}.
\newblock Accessed: 2025-03-15.

\bibitem{franklin2005databases}
M.~Franklin, A.~Halevy, and D.~Maier, ``From databases to dataspaces: a new abstraction for information management,'' {\em ACM Sigmod Record}, vol.~34, no.~4, pp.~27--33, 2005.

\bibitem{otto2022evolution}
B.~Otto, ``The evolution of data spaces,'' in {\em Designing data spaces: The ecosystem approach to competitive advantage}, pp.~3--15, Springer International Publishing Cham, 2022.

\bibitem{ids_2025}
I.~D.~S. Association, ``International data spaces.'' \url{https://internationaldataspaces.org/}.
\newblock Accessed: 2025-03-13.

\bibitem{nagel2021design}
L.~Nagel, J.~J. Hierro, E.~Perea, D.~Lycklama, C.~Mertens, A.-S. Taillandier, M.~Marques, J.~Gelhaar, A.~Marguglio, U.~Ahle, {\em et~al.}, ``Design principles for data spaces: position paper,'' tech. rep., E. ON Energy Research Center, 2021.

\bibitem{sharif2022eidas}
A.~Sharif, M.~Ranzi, R.~Carbone, G.~Sciarretta, F.~A. Marino, and S.~Ranise, ``The eidas regulation: a survey of technological trends for european electronic identity schemes,'' {\em Applied Sciences}, vol.~12, no.~24, p.~12679, 2022.

\bibitem{ids_ram_2025}
I.~D.~S. Association, ``Ids ram 4.0.'' \url{https://internationaldataspaces.org/}, 2022.
\newblock Accessed: 2025-03-13.

\bibitem{nagel2022build}
L.~Nagel and D.~Lycklama, ``How to build, run, and govern data spaces,'' in {\em Designing data spaces: The ecosystem approach to competitive advantage}, pp.~17--28, Springer International Publishing Cham, 2022.

\bibitem{sciore2020jdbc}
E.~Sciore and E.~Sciore, ``Jdbc,'' {\em Database Design and Implementation: Second Edition}, pp.~15--47, 2020.

\bibitem{aws_s3}
{Amazon Web Services}, ``{Amazon S3 - Cloud Object Storage}.'' \url{https://aws.amazon.com/s3/}.
\newblock Accessed: 2025-03-22.

\bibitem{durumeric2013analysis}
Z.~Durumeric, J.~Kasten, M.~Bailey, and J.~A. Halderman, ``Analysis of the https certificate ecosystem,'' in {\em Proceedings of the 2013 conference on Internet measurement conference}, pp.~291--304, 2013.

\bibitem{eclipse_edc_connector}
{Eclipse Foundation}, ``{Eclipse EDC Connector}.'' \url{https://github.com/eclipse-edc/Connector}.
\newblock Accessed: 2025-03-22.

\bibitem{fiware_true_connector}
{FIWARE Foundation}, ``{FIWARE True Connector}.'' \url{https://fiware-true-connector.readthedocs.io/en/latest/}.
\newblock Accessed: 2025-03-22.

\bibitem{eu_simpl}
{European Commission}, ``{Simpl: Cloud-to-edge federations empowering EU data spaces}.'' \url{https://digital-strategy.ec.europa.eu/en/policies/simpl}.
\newblock Accessed: 2025-03-22.

\bibitem{ianella2007open}
R.~Ianella, ``Open digital rights language (odrl),'' {\em Open Content Licensing: Cultivating the Creative Commons}, 2007.

\bibitem{IDSA2025}
{International Data Spaces Association}, ``{IDSA Infographic: Data Sharing in a Data Space}.'' \url{https://internationaldataspaces.org/wp-content/uploads/IDSA-Infographic-Data-Sharing-in-a-Data-Space.pdf}.
\newblock Accessed: 2025-03-30.

\bibitem{tardieu2022role}
H.~Tardieu, ``Role of gaia-x in the european data space ecosystem,'' in {\em Designing Data Spaces: The Ecosystem Approach to Competitive Advantage}, pp.~41--59, Springer International Publishing Cham, 2022.

\bibitem{dssc}
{Data Spaces Support Center}, ``{Data Spaces Support Center}.'' \url{https://dssc.eu/}.
\newblock Accessed: 2025-03-15.

\bibitem{dssc_blueprint}
{Data Spaces Support Center}, ``{Data Spaces Blueprint}.'' \url{https://dssc.eu/space/BPE/179175433/}, 2024.
\newblock Accessed: 2025-03-15.

\bibitem{dsba}
{Data Spaces Business Alliance}, ``{Data Spaces Business Alliance}.'' \url{https://data-spaces-business-alliance.eu/}.
\newblock Accessed: 2025-03-15.

\bibitem{bdva}
{Big Data Value Association}, ``{Big Data Value Association}.'' {\url{https://bdva.eu/}}.
\newblock Accessed: 2025-03-15.

\bibitem{fiware2025}
F.~Foundation, ``Fiware foundation.'' \url{https://www.fiware.org/foundation/}.
\newblock Accessed: 2025-03-13.

\bibitem{eclipse_edc}
{Eclipse Foundation}, ``{Eclipse Data Components (EDC)}.'' \url{https://projects.eclipse.org/projects/technology.edc}.
\newblock Accessed: 2025-03-22.

\bibitem{ids_dataspace_protocol}
{International Data Spaces Association}, ``{IDS Knowledge Base - Dataspace Protocol}.'' \url{https://docs.internationaldataspaces.org/ids-knowledgebase/dataspace-protoco}.
\newblock Accessed: 2025-03-22.

\bibitem{eona_x}
{EONA-X}, ``{EONA-X: The Tourism, Mobility and Logistics Data Space}.'' \url{https://eona-x.eu/}.
\newblock Accessed: 2025-03-22.

\bibitem{AWSDataSpaces}
{Amazon Web Services}, ``{Examples of data spaces built on top of AWS services}.'' \url{https://docs.aws.amazon.com/prescriptive-guidance/latest/strategy-building-data-spaces/data-spaces-on-aws-services.html}, 2024.
\newblock Accessed: 2025-03-15.

\bibitem{NTTData2025}
{NTT Data}, ``{The Power of Data Spaces}.'' \url{https://benelux.nttdata.com/insights/whitepapers/the-power-of-data-spaces}, 2024.
\newblock Accessed: 2025-03-15.

\bibitem{desemantic}
A.~De~Craene, Z.~Vanlishout, and P.~Colpaert, ``Semantic and technically interoperable data exchange in the flanders smart data space,''

\bibitem{lush2024assessing}
V.~Lush, L.~Bastin, K.~Otsu, and J.~Mas{\'o}, ``Assessing fairness of citizen science data in the context of the green deal data space,'' {\em International Journal of Digital Earth}, vol.~17, no.~1, p.~2344587, 2024.

\bibitem{wilkinson2016fair}
M.~D. Wilkinson, M.~Dumontier, I.~J. Aalbersberg, G.~Appleton, M.~Axton, A.~Baak, N.~Blomberg, J.-W. Boiten, L.~B. da~Silva~Santos, P.~E. Bourne, {\em et~al.}, ``The fair guiding principles for scientific data management and stewardship,'' {\em Scientific data}, vol.~3, no.~1, pp.~1--9, 2016.

\bibitem{etip2023}
E.~SNET, ``Energy data space – policy paper.'' \url{https://horizoneuropencpportal.eu/sites/default/files/2024-05/etip-snet-energy-data-space-policy-paper-2023.pdf}, 2023.
\newblock Accessed: 2025-05-09.

\bibitem{bridge2021}
E.~Commission, ``Bridge initiative: Coordinating energy research and innovation projects across the eu,'' 2021.
\newblock Accessed: 2025-05-09.

\bibitem{omegaXWebsite}
{OMEGA-X Consortium}, ``Omega-x project.'' https://omega-x.eu/.
\newblock Accessed: 2025-05-09.

\bibitem{enershareWebsite}
{ENERSHARE Consortium}, ``Enershare project.'' https://enershare.eu/.
\newblock Accessed: 2025-05-09.

\bibitem{omega2025standardization}
A.~Kung, ``Omega-x standardisation workshop – presentation slides.'' https://omega-x.eu/wp-content/uploads/2025/04/Omega-X-Standardisation-workshop-Antonio-Kung.pdf, 2025.
\newblock Accessed: 2025-05-09.

\end{thebibliography}
\end{document}